\documentclass[12pt]{article}

\usepackage{amsmath,amssymb,amssymb,graphicx} %use drftcite in draftmode
\usepackage{cite}
\newcommand{\dosum}[2]{\sum_{\mathrel{\rlap{\lower6pt\hbox{\scriptsize{\mbox{$#2$}}}}\raise2pt\hbox{\scriptsize{\mbox{$#1$}}}}}}

\newcommand{\beq}{\begin{eqnarray}}% can be used as {equation} or  {eqnarray}
\newcommand{\eeq}{\end{eqnarray}}

\newcommand{\centeron}[2]{{\setbox0=\hbox{#1}\setbox1=\hbox{#2}\ifdim
\wd1>\wd0\kern.5\wd1\kern-.5\wd0\fi \copy0
\kern-.5\wd0\kern-.5\wd1\copy1\ifdim\wd0>\wd1
                                   \kern.5\wd0\kern-.5\wd1\fi}}
\newcommand{\ltap}{\>\centeron{\raise.35ex\hbox{$<$}}
                           {\lower.65ex\hbox{$\sim$}}\>}
\newcommand{\gtap}{\>\centeron{\raise.35ex\hbox{$>$}}
                           {\lower.65ex\hbox{$\sim$}}\>}

\newcommand{\lsim}{\mathrel{\ltap}}

\newcommand\ZZ{\hbox{\zfont Z\kern-.4emZ}}
\font\zfont = cmss10 %scaled \magstep1

\newcommand{\drawsquare}[2]{\hbox{%
\rule{#2pt}{#1pt}\hskip-#2pt%  left vertical
\rule{#1pt}{#2pt}\hskip-#1pt%  lower horizontal
\rule[#1pt]{#1pt}{#2pt}}\rule[#1pt]{#2pt}{#2pt}\hskip-#2pt%  upper horizontal
\rule{#2pt}{#1pt}}% right vertical

\newcommand{\fund}{\drawsquare{6.5}{0.4}}%  fundamental

\newcommand{\asymm}{\raisebox{-3.5pt}{\drawsquare{6.5}{0.4}\hskip-6.9pt%
        \raisebox{6.5pt}{\drawsquare{6.5}{0.4}}}}%  antisymmetric second rank

\begin{document}
\begin{titlepage}

\vskip.5cm
\begin{center}
{\huge \bf A Seiberg Dual for the MSSM:}\\
\vskip 8pt
{\huge \bf Partially Composite $W$ and $Z$}

\vskip.1cm
\end{center}
\vskip0.2cm

\begin{center}
{\bf
{Csaba Cs\'aki}$^{a}$, {Yuri Shirman}$^{b}$
{\rm and}
{John Terning}$^{c}$}
\end{center}
\vskip 8pt

\begin{center}
$^{a}$ {\it  Department of Physics, LEPP, 
Cornell University, Ithaca, NY 14853} \\
\vspace*{0.1cm}
$^{b}$ {\it
Department of Physics, University of California, Irvine, CA
92617} \\
\vspace*{0.1cm}
$^{c}$ {\it
Department of Physics, University of California, Davis, CA
95616} \\

\vspace*{0.3cm}
{\tt  csaki@cornell.edu, yshirman@uci.edu, terning@physics.ucdavis.edu}
\end{center}

\vglue 0.3truecm

\begin{abstract}
\vskip 3pt
\noindent
We examine the possibility that the $SU(2)_L$ gauge group of the standard model appears as the dual ``magnetic" gauge group
of a supersymmetric gauge theory, thus the $W$ and $Z$ (and through mixing, the photon) are composite (or partially composite) gauge bosons.  Fully composite gauge bosons are expected to interact strongly at the duality scale, and a large running is needed to match the electroweak gauge couplings. Alternatively one can mix the composite ``magnetic" gauge bosons with some elementary ones to obtain realistic models. In the simplest and most compelling example the Higgs and top are composite, the $W$ and $Z$ partially composite and the light fermions elementary. The effective theory is an NMSSM-type model where the singlet is a component of the composite meson. There is no little hierarchy problem and the Higgs mass can be as large as 400 GeV. This ``fat Higgs"-like model can be considered as an explicit 4D  implementation of RS-type models with gauge fields in the bulk. 

\end{abstract}

\end{titlepage}

%\renewcommand{\thefootnote}{(\arabic{footnote})}

%%%%%%%%%%%%%%%%%%%%%%%%%%%%%%%%%%%%%%%%%%%%%%%%%%%%%%
%%%%%%%%%%%%%%%%%%%%%%%%%%%%%%%%%%%%%%%%%%%%%%%%%%%%%%
\section{Introduction}
\label{sec:intro}
\setcounter{equation}{0}
\setcounter{footnote}{0}
%%%%%%%%%%%%%%%%%%%%%%%%%%%%%%%%%%%%%%%%%%%%%%%%%%%%%%
%%%%%%%%%%%%%%%%%%%%%%%%%%%%%%%%%%%%%%%%%%%%%%%%%%%%%%

Long ago Abbott and Farhi \cite{Abbott} showed (via complementarity \cite{Banks}) how in the strong coupling limit of standard model (SM) weak interactions there would still be massive, composite, spin-1 particles with the quantum numbers of the $W$ and $Z$.  This was a startling and intriguing idea, but there was no explanation for why these particles would couple like the ordinary $W$ and  $Z$ gauge bosons.

After the discovery of Seiberg duality \cite{Seiberg} and the existence of massless composite dual ``magnetic" gauge fields the speculation that the $W$ and $Z$ could themselves be ``magnetic" composites emerging at low energies was revived. Seiberg himself commented on this possibility, and there have been several early  attempts \cite{falsstarts} along these lines.  The goal of identifying the $W$ and $Z$ as ``magnetic" composites became even more appealing recently due to Komargodski's (re)-interpretation 
\cite{Zohar} of Seiberg duality, where the ``magnetic" gauge fields are identified as the $\rho$-mesons of a strongly interacting electric theory. Komargodski also clarified the relation between Seiberg duality on the one-hand and the ideas of the Georgi vector-limit \cite{Georgi} and hidden-local symmetry~\cite{hidden} on the other. He was able to show that the former is a better fit to QCD phenomenology than the latter, giving some hope that Seiberg's ideas can be extended to the non-supersymmetric regime.

Similar ideas about composite SM gauge fields have also been put forward in the context of warped extra dimensional theories~\cite{Tony1,Tony2}. In fact the original Randall-Sundrum (RS1) model \cite{RS} itself can be interpreted as a composite SM using the AdS/CFT dictionary. However, in these models it is also not clear why the composite gauge fields would be light or posses a real gauge invariance dictating the interactions with  the matter fields and themselves. 

A more plausible version of the RS1 model is the so called ``realistic RS1" \cite{realisticRS} with the gauge fields in the bulk and SM fermions mostly on the UV brane (except the third generation quarks). The interpretation of this model is that the gauge fields are mixtures of elementary and composite gauge bosons, thus a weakly coupled $SU(2)_L\times U(1)_Y$ is possible. Similarly the SM fermions are mostly elementary, while the Higgs and the top is fully  composite. 

In this paper we present various supersymmetric ``electric" theories leading to the MSSM at low energies. The ``electric" theory is below (or just at the edge of) the conformal window, thus the 
emergence of composite dual ``magnetic" gauge fields is expected. However, at the matching scale 
the ``magnetic'' description is expected to be strongly coupled unlike the weak interactions of the SM.
To lower the coupling one needs a larger flavor symmetry in the electric description --- this both lowers the ``magnetic'' coupling at the matching scale and accelerates the RG evolution.
Alternatively, the coupling may be reduced via mixing with elementary gauge bosons like in the realistic RS1 models. 
While discussing both of these possibilities we will concentrate on the latter approach:
we assume that in addition to the composite $SU(2)$ there is also an $SU(2)$ subgroup of the global symmetries that is gauged, and a VEV for the dual ``quarks" breaks these to the diagonal subgroup, producing the weakly coupled $W$ and the $Z$. Below this VEV the low energy theory will in fact be a next-to-MSSM (NMSSM) type model with composite singlet, Higgs, and top fields.
The nature of the light fermions depends on specific realization of the model.
We will present two constructions. In the first, all the MSSM $SU(2)_L$ doublets arise
as dual ``magnetic quarks'', while the $SU(2)_L$ singlets arise as mesons. The addition of elementary versions of the right-handed SM fermions that mix with their composite relatives will allow a realistic Yukawa structure to be obtained.
In the second variant of the model all the light fermions are elementary, which corresponds to a supersymmetric RS-type model like~\cite{Yasunori}. They can acquire mass through coupling to an elementary Higgs fields that, in turn, couples to composite Higgses in an analogy with the construction of \cite{fathiggs1,fathiggs2} (see also \cite{Kitano}).
The new feature of our models is that in addition to Higgses the gauge fields as well as (some of) the SM fermions are also partially composite (or ``fat''), similar to \cite{fathiggs3}.

The paper is organized as follows. In section \ref{sec:composite?} we argue that one would not expect to be able to find a model with fully composite $W$ and $Z$ that can reproduce the properties of the SM. We discuss a fully composite toy model in section \ref{sec:toymodel}. In section \ref{sec:lowscales} we discuss modifications necessary to make the model realistic. Of special interest is the scenario where the compositeness scale is low. We show that this can be achieved if composite gauge bosons are mixed with elementary ones. As a result the $W$ and $Z$ are only partially composite. In section \ref{sec:minimalpartial} we introduce the minimal model where light SM fermions are elementary, the top and Higgs are composite, while the $W$ and $Z$ are partially composite. We summarize our results in section \ref{sec:conclusions}.

 %%%%%%%%%%%%%%%%%
 %%%%%%%%%%%%%%%%%
 \section{Composite $W$ and $Z$?}
\label{sec:composite?}
\setcounter{equation}{0}
\setcounter{footnote}{0}
%%%%%%%%%%%%%%%%%%%%%%%%%%%%%%%%%%%%%%%%%%%%%%%%%%%%%%%%%%%%%%

In this section we argue that the properties of the SM $W$ and $Z$ are not generically expected to be reproduced in a fully composite model, like the Abbott-Farhi model, the original RS1 or a straight Seiberg duality.\footnote{We thank Markus Luty for focusing our attention on this issue.} The gauge coupling of the $SU(2)_L$ at the weak scale is $g\sim 0.65$. However, the coupling of a composite gauge boson at the compositeness scale is expected to be strong. According to Naive Dimensional Analysis (NDA)~\cite{NDA} expectations one should have $g \sim 4 \pi/\sqrt{N}$. Phenomenologically for the $\rho$ of QCD one finds $g \sim 6 \sim 4 \pi/\sqrt{3}$. Thus one would need a very large logarithmic running in order to reduce the coupling to the observed level. 

As an example consider the case of interest involving Seiberg duality (this case was also explored in~\cite{Tony2}). Here ``magnetic" gauge bosons will play the role of the $W$ and $Z$ while ``magnetic quarks" will correspond to $SU(2)_L$ doublets. 
Thus we are considering an $SU(N)$ ``electric'' theory with $F=N+2$ flavors although the following discussion is more general.
For general $F$ and $N$, the matching of dynamical scales (strong interaction scale, or in the case of IR free dual, Landau pole)
is given by~\cite{Seiberg} 
\begin{equation}
\Lambda_{\rm el}^{b_{\rm el}} \Lambda_{\rm mag}^{b_{\rm mag}} = (-1)^N \Lambda^{b_{\rm el}+b_{\rm mag}}~,
\label{matching}
\end{equation}
where $\Lambda_{\rm el, mag}$ are the ``electric" and ``magnetic" dynamical scales 
and $b_{\rm el}= 3N-F$, $b_{\rm mag}=3(F-N)-F$ are the one-loop $\beta$-function coefficients.  We can rewrite Eq. (\ref{matching}) as
\beq
\frac{1}{g_{\rm el}^2(|\Lambda|)}= \frac{b_{\rm el}}{8 \pi^2} \log \left( \frac{|\Lambda|}{\Lambda_{\rm el}} \right) = -\frac{b_{\rm mag}}{8 \pi^2} \log \left( \frac{|\Lambda|}{\Lambda_{\rm mag}} \right) =-\frac{1}{g_{\rm mag}^2(|\Lambda|)}~.
\eeq

\begin{figure}[htb]
\begin{center}
\includegraphics[width=7cm]{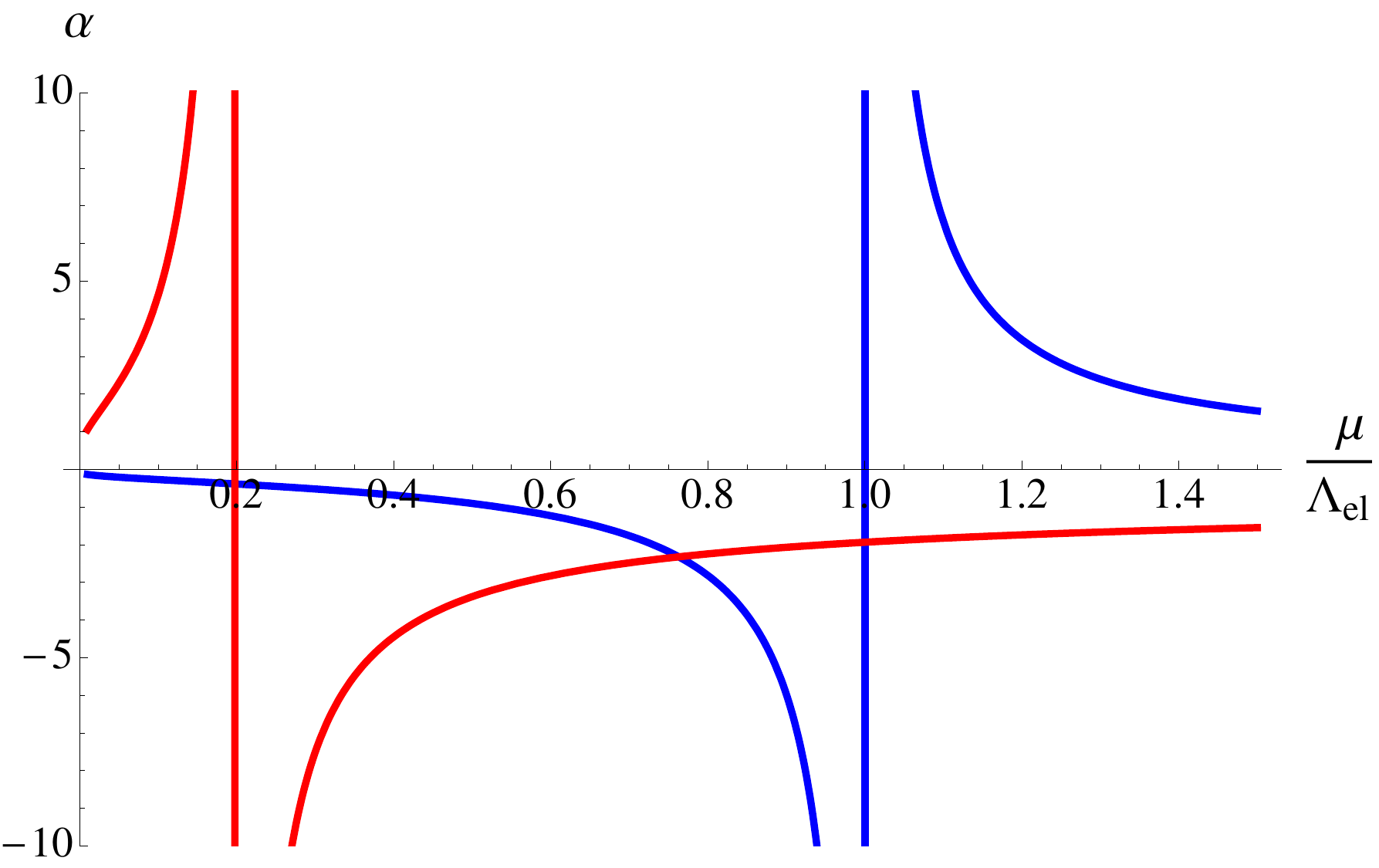}\hspace*{1cm} \includegraphics[width=7cm]{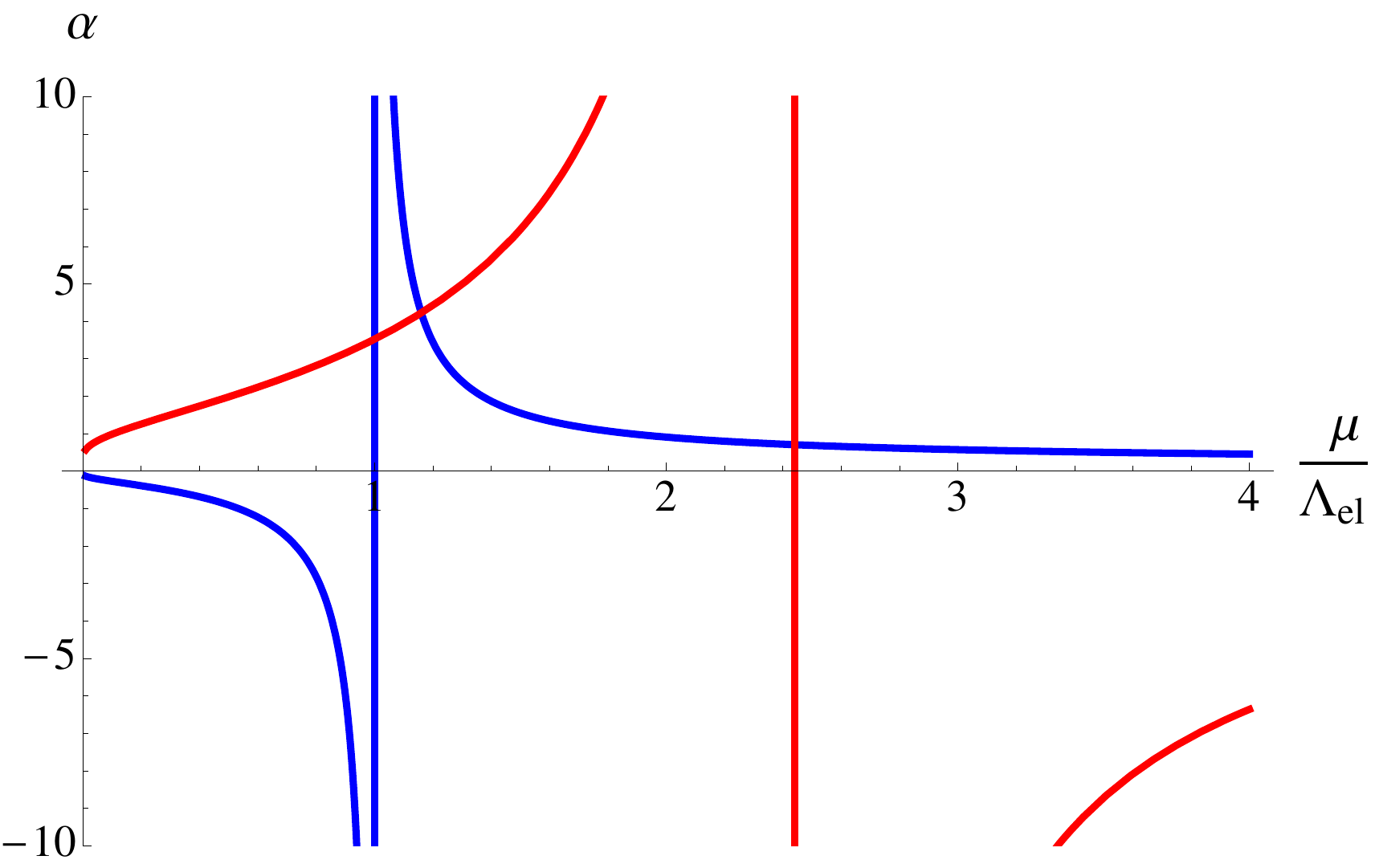}
\end{center}
\caption{Values of the one-loop gauge couplings $\alpha =g^2/4 \pi$, as functions of the renormalization scale $\mu$, for $\Lambda = 1.5\, \Lambda_{\rm el}$ on the left and  $\Lambda = 0.8\, \Lambda_{\rm el}$ on the right.  The ``electric" coupling (shaded in blue for 6 colors and 8 flavors) is positive for $\mu > \Lambda_{\rm el}$, while the ``magnetic'' coupling (shaded in red) is positive for $\mu < \Lambda_{\rm mag}$.}
 \label{fig:couplings}
\end{figure}

We see that formally $|\Lambda|$ is the scale where the two dual couplings are equal up to a sign, although  this means  (outside of the conformal window) that we have extrapolated the coupling of one of the theories  to a scale, $|\Lambda|$, that is beyond its range of validity. 
For example, if  we choose $|\Lambda|>\Lambda_{\rm el}$ for a theory in a free ``magnetic" phase, we see that the ``magnetic'' theory is renormalized at a scale $|\Lambda|$ that is above its Landau pole.
In this case there is a gap between $\Lambda_{\rm mag}$  and $\Lambda_{\rm el}$  and a simple description of the dynamics is not known.  This situation is shown on the left in Fig. \ref{fig:couplings}.  
If $|\Lambda|$ is smaller than $\Lambda_{\rm el}$ then the Landau pole of the ``magnetic'' theory is above $\Lambda_{\rm el}$, as shown on the right in  Fig. \ref{fig:couplings}. 
The ambiguity in the value of $|\Lambda|$ is irrelevant for  Seiberg duality: since the duality holds in the extreme infrared, for a fixed ``electric'' theory, any value of $|\Lambda|$ will lead to the same ``magnetic" dynamics sufficiently far in the infrared.

Given a particular underlying  ``electric" theory that is valid above $\Lambda_{\rm el}$ we can run into at least two problems when we try to extend Seiberg duality for a free ``magnetic" phase beyond the infrared. First there will be unknown, heavy, composite states, that are not included in the ``magnetic" description. However, as long as these states have masses around
$\Lambda_{\rm el}$, then  as far as the low-energy theory goes their effects can be ``absorbed" into threshold corrections at the matching scale. The second, more serious,  problem is that there should be a specific value of $|\Lambda|$ which gives the best description of physics in the weakly coupled ``magnetic" theory as the renormalization scale is raised.
Currently it is not known how to fix the correct value of $|\Lambda|$ for Seiberg duality. NDA suggests that we should choose $|\Lambda|\sim\Lambda_{\rm el}\sim\Lambda_{\rm mag}$.  

If it is possible to select a theory where  the correct value of $|\Lambda|$ is much smaller than  $\Lambda_{\rm el}$ we would have a ``magnetic" theory that has a weak gauge coupling at the scale where it matches on to the underlying strong dynamics.
How can this be possible?  Recall that the physics of the ``magnetic" phase is parameterized by two coupling constants, the gauge coupling and the dynamical Yukawa coupling of the ``meson"  to the dual ``quarks". Both of these couplings have a Landau pole. Since even at one-loop the gauge coupling appears in the Yukawa $\beta$-function, it is quite plausible that at strong gauge coupling the two Landau poles are closely related. Nevertheless at weak gauge coupling  there is no definitive argument (other than NDA) why they should both be at the same scale.  This suggests that $|\Lambda| < \Lambda_{\rm el}$ describes the case where the Landau pole, $\Lambda_Y$,  of the Yukawa coupling is below the Landau pole of the gauge coupling:  $\Lambda_Y< \Lambda_{\rm mag}$. This conjecture is supported by the fact that the tree-level value of the Yukawa coupling, given by the dual superpotential, contains  a factor $\Lambda_{\rm el}/|\Lambda|$.  So as $|\Lambda|$ is lowered, the Yukawa coupling is enhanced by an inverse factor of $|\Lambda|$, while the ``magnetic'' gauge coupling at the matching scale is suppressed by $\log \Lambda_{\rm el}/|\Lambda|$.  More explicitly
\beq
\frac{1}{g_{\rm mag}^2(\Lambda_{\rm el} )}= \frac{b_{\rm mag}}{8 \pi^2} \log \left( \frac{\Lambda_{\rm el} }{\Lambda_{\rm mag}} \right)= \frac{b_{\rm el}+b_{\rm mag}}{8 \pi^2} \log  \frac{\Lambda_{\rm el}}{|\Lambda|} = \frac{F}{8 \pi^2} \log  \frac{\Lambda_{\rm el}}{|\Lambda|} ~.
\eeq
If we assume that the Landau pole for the Yukawa coupling of the ``magnetic'' description can not be much lower than the strong coupling scale of the ``electric'' theory, then  this provides  a lower bound on $|\Lambda|$.
On the other hand, we are interested in the models where $SU(2)_L$ doublets arise as ``magnetic quarks'' of the dual theory.  Since at the weak scale the top Yukawa is $\mathcal{O}(1)$  and, in NMSSM-like scenarios, the Higgs coupling to a singlet is also required to be $\mathcal{O}(1)$, so we also have  an upper bound on $|\Lambda|$. Thus the scale $|\Lambda|$ is bounded by
\begin{equation}
 1\lsim y\approx\frac{\Lambda_{\rm el}}{|\Lambda|}\lsim 4\pi\,.
\end{equation}

At the same time, we are interested in the smallest possible value of the ``magnetic" coupling. Choosing $\Lambda_{\rm el}/|\Lambda|=4\pi$ we find
\beq
g_{\rm mag}^2(\Lambda_{\rm el} )=\frac{8 \pi^2} {F \log \left( 4 \pi \right) } \approx \frac{31}{F}~.
\label{magneticmatching}
\eeq
Given the number of light $SU(2)_L$ doublets in the ``magnetic'' description (using the one-loop $\beta$-function coefficient $b_m$) we may relate $\Lambda_{\rm el}$ to the weak scale. If $b_m$ is small (as is the case in MSSM) we find $\Lambda_{\rm el}\gg M_{\rm Pl}$.
Thus minimal models of this type are problematic and a fully composite $W$ and $Z$ are difficult to implement in a realistic theory.
Nevertheless, in the following section we present a minimal module that would lead to a fully composite MSSM, which is generically expected to be strongly interacting at the duality scale. There are two possibilities to remedy this problem. First, one could enlarge the flavor symmetry of the ``electric'' description. This reduces $g_{\rm mag}^2$ at the matching scale. Furthermore, larger values of $F$ correspond to more doublets in the ``magnetic'' theory. If the doublets are light (they can be given vector-like masses anywhere between the weak scale and $\Lambda_{\rm el}$) faster RG running allows for a lower $\Lambda_{\rm el}$. Alternatively, one may mix the ``magnetic'' gauge bosons with some elementary fields --- in which case the $W$ and $Z$ would no longer be fully composite. We will explore both of these possibilities in the sections \ref{sec:lowscales} and \ref{sec:minimalpartial} and see that the most interesting results are obtained when a combination of two approaches is used.

 %%%%%%%%%%%%%%%%%%%%%%%%%%%%%%%%%%%%%%%%%%%%%%%%%%%%%%%%%%%%%%
\section{The minimal module: a fully composite Seiberg dual for the MSSM}
\label{sec:toymodel}
\setcounter{equation}{0}
\setcounter{footnote}{0}
%%%%%%%%%%%%%%%
%%%%%%%%%%%%%%%

Our first aim is to find a  Seiberg duality with fully composite $W$, $Z$ bosons and also fully composite SM fermions. 
We want the  $SU(2)_L$ in this theory to be the low-energy description of the free ``magnetic" phase of the ``electric" theory. If we restrict ourselves to the simplest SUSY QCD, $SU(N)$   with $F$ flavors, then we obviously need $F-N=2$. Since the $SU(2)_L$  is now a composite gauge group, all $SU(2)_L$ doublets should also appear as composites. The only candidates for such doublets are the dual ``magnetic quarks" $q,\bar{q}$. In the MSSM there are 9 quarks, 3 leptons and 2 Higgs doublets. This implies $F\geq 7$. It turns out that in order to generate all the Yukawa couplings for the matter fields one actually needs two sets of Higgses $H_{u,d}$ and $H_{u,d}'$. Thus $N=6, F=8$ model is the smallest possible ``electric" theory sufficient to incorporate all MSSM matter fields and Yukawa couplings. 

The ``electric" theory is 
\beq	
\begin{array}{c|c|cccc}
& SU(6) & SU(8)_1  & SU(8)_2 & U(1)_V & U(1)_R \\
\hline
{\cal Q}& \fund & \overline{\fund}   & {\bf 1} & \frac{1}{24}  & {\frac{1}{4}}
\vphantom{\raisebox{3pt}{\asymm}}\\
{\cal \bar Q} & \overline{\fund}   & {\bf 1}  & \overline{\fund}
& -\frac{1}{24} & {\frac{1}{4}} \vphantom{\raisebox{3pt}{\asymm}}\\
\end{array}
\label{Chap3:gauge:SUSYQCD}
\eeq
where the $SU(6)$ is the gauge group while the other groups are the global symmetries.  The usual 
dual theory is
\beq	
\begin{array}{c|c|cccc}
& SU(2)_L & SU(8)_1  & SU(8)_2 & U(1)_V & U(1)_R \\
\hline
q& \fund & \fund & {\bf 1} & \frac{1}{8}  & {\frac{1}{4}}
\vphantom{\raisebox{3pt}{\asymm}}\\
\bar q& \overline{\fund}   & {\bf 1}  & \fund
& - \frac{1}{8} & {\frac{1}{4}} \vphantom{\raisebox{3pt}{\asymm}}\\
M& {\bf 1} &  \overline{\fund}   &  \overline{ \fund} & 0 & {\frac{3}{2}}
\vphantom{\raisebox{3pt}{\asymm}}\\
\end{array}
\label{Chap3:gauge:SUSYdual}
\eeq
with the superpotential
\beq
W= y \bar q M q~.
\eeq
As explained earlier the $SU(2)_L$ gauge bosons emerge at low energies as massless ``magnetic" composite gauge bosons. The remaining gauge symmetries of the SM,  $SU(3)_{c}\times U(1)_Y$, will be embedded into the global symmetries.

In order to be able to identify the SM fermions with the dual fields one needs to decompose  
the flavor symmetry groups in the following way:
\begin{equation}
\begin{split}
&SU(8)_1 \supset SU(3) \times SU(3) \times SU(2)_{R,1} \\
&SU(8)_2 \supset SU(3)_G \times SU(3) \times SU(2)_{R,2} 
\end{split}
\end{equation}
where $SU(3)_G$, is a generational symmetry for the leptons, and $SU(3)_c$ gauges the diagonal subgroup of the three other $SU(3)$'s.  This embedding of $SU(3)_c$ into the global symmetry of $SU(6)$ SUSY QCD is anomalous by itself, and we will shortly introduce elementary spectator fields to cancel anomalies.

The dual ``quarks" can be identified with the SM doublets as follows:
\begin{equation}
\begin{split}
&q= \left(\begin{array}{c} t_n \\ b_n \end{array}\right)_L , \left(\begin{array}{c} c_n \\ s_n \end{array}\right)_L , H_u , H_d^\prime \\
&\bar q= \left(\begin{array}{c} \nu_e \\ e\end{array}\right)_L , \left(\begin{array}{c} \nu_\mu \\ \mu \end{array}\right)_L ,  \left(\begin{array}{c} \nu_\tau \\ \tau \end{array}\right)_L , \left(\begin{array}{c} u_n \\ d_n \end{array}\right)_L , H_d , H_u^\prime 
\end{split}
\end{equation}
where the color index $n=1,2,3$. The meson decomposes as 
\beq
M=\left(\begin{array}{ccc}  V^{1,j}_{n} & V^{2,j}_{n}  & \begin{array}{c} \bar \nu_e\,\, \bar e  \\ \bar \nu_\mu\,\, \bar \mu  \\ \bar \nu_\tau\,\, \bar \tau\end{array} \\
C^1_{p}\,\epsilon_{mnp}+X^1_{m,n} & C^2_{p}\,\epsilon_{mnp}+X^2_{m,n}  \vphantom{\sqrt{\frac{F^j}{g_g}}} & \bar u_n\,\, \bar d_n  \\
\begin{array}{c} \bar b_n \\ \bar t_n \end{array} & \begin{array}{c} \bar s_n \\ \bar c_n \end{array}  & \begin{array}{cc} S & T^- \\ T^+ & S' \end{array}  \end{array} \right)
\eeq
where $j$ is a generation index and $V^{i,j}_{n} $ are 3 $\times$ 3 matrices representing three generations of color anti-triplets, $C^i_{n}$ are  color triplets, while $X^i_{m,n}$  are 3 $\times$ 3 symmetric matrices 
representing  conjugate sextets. $S$, $S'$, and  $T$ are color and $SU(2)_L$ singlets. 
In a phenomenologically viable theory these composite spectators must be heavy. Thus we add elementary fields, $\bar V^{1,2}$, $\bar C^{1,2}$, and $\bar X^{1,2}$ in conjugate representations so that all the spectators are vector-like. The mass of the spectator fields can arise from dimension three superpotential terms in the ``electric'' description and therefore is generally unsuppressed.
Incidentally we note that the addition of spectators cancels all anomalies since  in the ``magnetic" description all the fields that are not part of the MSSM  are  vector-like. The consequence of these many additional fields will be that the QCD $\beta$ function will be very large (and not asymptotically free). With the smallest matter content outlined above we find $b_{QCD}=14$ implying a Landau pole for QCD within a couple of decades of running above the duality scale. In this case a duality cascade of the sort envisioned in~\cite{Strassler} and~\cite{KlebanovStrassler} will be inevitable. 

Hypercharge can be identified as a subgroup of the anomaly free (non-R) global $U(1)$ symmetries. These include besides the $U(1)_V$ the diagonal generators of $SU(8)_{1,2}$ which have not been gauged and are generation independent. These generators are:
\begin{equation}
\begin{split}
& T^3_{(i)} = \left( \begin{array}{cccc} {\bf I} &&&\\  & -{\bf I} && \\ &&  0 & \\ &&& 0 \end{array} \right)~, \quad \quad 
T^8_{(i)} = \left( \begin{array}{cccc} {\bf I} &&&\\  & {\bf I} && \\ &&  -3 & \\ &&& -3 \end{array} \right) \\
& T^{R,3}_{(i)} = \left( \begin{array}{cccc} 0&&&\\  & 0 && \\ &&  \frac{1}{2}  & \\ &&& -\frac{1}{2}  \end{array} \right)
\end{split}
\end{equation}
where ${\bf I} $ is the 3 $\times$ 3 identity matrix.  Then hypercharge can be written as
\beq
Y= Q_V +\left(T^{R,3}_{(1)} -T^{R,3}_{(2)} \right) +\frac{1}{24} \left(T^8_{(1)} -T^8_{(2)}  \right) -\frac{1}{3} \,T^3_{(2)} 
\eeq
This choice will ensure that all SM fields have the correct hypercharge quantum number, while the hypercharge of the additional fields is $+1/3$ for $V^{1,2}$ and $-1/3$ for $C^{1,2},X^{1,2}$. 
 
Even though this model contains ``leptoquarks" (the $V^{1,2}$ fields in the composite meson $M$) 
we can easily see that both baryon and lepton number is preserved in this model. These symmetries don't have to be anomaly free (as they are not in the SM or the MSSM), and thus do not have to be a subgroup of the anomaly free global symmetries. The obvious charge assignments are
\begin{eqnarray} 
&B_q = {\rm diag} (\frac{1}{3}, \frac{1}{3}, \frac{1}{3}, \frac{1}{3}, \frac{1}{3}, \frac{1}{3}, 0,0), \nonumber  \ \ &L_q=0\nonumber  \\ 
&B_{\bar{q}}=  {\rm diag} (0,0,0, \frac{1}{3}, \frac{1}{3}, \frac{1}{3}, 0,0), \ \ &L_{\bar{q}}=  {\rm diag} (1,1,1, 0,0,0, 0,0)
\end{eqnarray}
while the charge assignment for the meson $M$ is such that $\bar{q}Mq$ is invariant: the right handed anti-quarks carry baryon number $-1/3$ and no lepton number, the right handed leptons carry lepton number -1 and no baryon number, the singlets $S$, $S'$, and $T$ carry no lepton or baryon number. The charge assignments for the ``leptoquarks" are 
$V^{1,2}: (-1/3, -1)$, and for the ``diquarks" $C^{1,2}, X^{1,2}: (-2/3, 0)$. By construction all interaction terms including those of $V,C,X$ conserve lepton number. 

This model will essentially be an NMSSM-type theory with two sets of Higgs doublets. We will also assume that a tree-level mass term in the ``electric" theory is present which will map to tadpole superpotential terms, $f^2 S+f'^2S'$, in the dual. We will also add soft supersymmetry breaking scalar masses and eventually also an $A$-term for the cubic interaction. The $A$-term will be the sources of the $S,S'$ VEVs, and thus the effective $\mu$-terms giving rise to Higgsino masses. Since the Higgs quartic is mostly from the NMSSM-like coupling there will be no relation to the $Z$ mass and obviously there will be no little hierarchy problem. 

The Yukawa interactions from $\bar{q} M q$ involving the fields that carry the quantum numbers of the SM fermions are given by
\begin{eqnarray}
\label{fermionYukawas}
y \left[ L_i H_u \bar{\nu}_i + L_i H_d' \bar{e}_i + Q_1 H_u \bar{u}_1 + Q_1 H_d' \bar{d}_1 +
 Q_j H_d \bar{d}_j + Q_j H_u' \bar{u}_j  \right]~,
\end{eqnarray}
where $i=1,2,3$ and $j=2,3$.
Since all of the fields are composite, each field (as expected) will have an ${\cal O}(1)$ Yukawa coupling involving one of the four Higgs doublets. While QCD and hypercharge interactions break the flavor symmetries of the ``electric" theory, the Yukawa's are still diagonal in the flavor space. In order to reduce the Yukawa couplings to their SM values and reproduce the correct flavor structure of mass matrices one needs to introduce flavor physics through a set of elementary vector-like fields for every light SM multiplet. The new fields will mix with RH composites making them heavy and effectively replacing them with elementary RH fields in the spectrum.
The mixing angle will suppress the effective Yukawa coupling giving rise to the correct SM mass fermion mass spectrum. Since the top is heavy, there is no need to suppress the top Yukawa coupling, so the RH top can remain composite. This type of fermion sector should be familiar from realistic RS models.

One interesting consequence of this model is that (part of) R-parity is an accidental low-energy symmetry.  
In this setup the operator $L^2$ is mapped to a baryon of the ``electric" theory, which is of the form ${\cal Q}_{\rm el}^6$, while $\bar{e}$ is a meson.  For example the $LLe$ operator is expected to be suppressed by $(\Lambda_{\rm el} /\Lambda_{UV})^5$, where $\Lambda_{UV}$ is a scale of UV completion of the ``electric" theory.

%%%%%%%%%%%%%%%%%%%%%%
%%%%%%%%%%%%%%%%%%%%%%
\section{Low Compositeness Scales: Partially composite $W$ and $Z$ and SM fermions}
\label{sec:lowscales}
\setcounter{equation}{0}
\setcounter{footnote}{0}
%%%%%%%%%%%%%%%%%%%%%%
%%%%%%%%%%%%%%%%%%%%%%

We have presented a Seiberg dual in the previous section that formally reproduces all the properties of the MSSM with all fields composite. However the generic expectation is that 
this model is strongly coupled. This is like the Abbott-Farhi (strongly coupled SM), or like the RS1 model. For example, if we look at 
  the ``magnetic" coupling constant at the duality scale (\ref{magneticmatching}) we find for $F=8$ the value of the coupling $g^2_{\rm mag}(\Lambda_{\rm el})\sim 3.9$. Assuming a running with the ``magnetic" $\beta$ function below this scale we find 
  \begin{equation}
  \Lambda_{\rm el} = M_{EW} e^{\frac{8\pi^2}{6-F} \left( \frac{1}{g^2 (M_{EW})} -\frac{F}{31} \right)}
  \end{equation}
 which for $F=8$ gives $\Lambda_{\rm el} \gg M_{Pl}$.  
 To remedy this problem we could increase the size of electric gauge group and the flavor symmetries keeping $F-N=2$ fixed. Such a modification adds several electroweak doublets as well as color triplets. Furthermore, additional elementary anti-triplets are required to maintain anomaly freedom as well as the vector-like nature of the spectator degrees of freedom. 
 Spectator quarks can be removed from the spectrum by adding tree-level dimension 3 superpotential operators in the ``electric'' theory that couple colored mesons to elementary spectator quarks.
On the other hand, to remove doublets one could introduce spontaneous breaking of the  ``electric'' gauge group down to $SU(6)$. In the ``magnetic" description this breaking will lead to VEVs for some of the mesons as well as mass for unwanted doublets.
 In general spectator masses can be tuned to any value between the electroweak scale and $\Lambda_{\rm el}$. It is preferable to give color triplets masses of order $\Lambda_{\rm el}$ to avoid an $SU(3)_c$ Landau pole. On the other hand, relatively light doublets allow one to lower $\Lambda_{\rm el}$ and keep $F$ as small as possible. 

As an example, an electric theory based on an $SU(9)$ gauge group with $F=11$ flavors can have composite $W$'s and $Z$'s with $\Lambda_{\rm el}\lsim 10^{15}\,{\rm GeV}$ if the  masses of the
additional light doublets is of order of a TeV. Of course, with such a high strong coupling scale the composite degrees of freedom will behave as elementary particles at LHC energies. If one insisted on a low compositeness scale of order 10 TeV, one would need to choose $F\sim 29$. In this case the running of the $SU(2)_L$ group will be very fast and a weakly coupled $W$ and $Z$ can be achieved. However now one might start worrying about the presence of all the light doublets with masses right around the electroweak scale ($\sim 100$ GeV). 

A more reasonable approach is to  mix the ``magnetic" $SU(2)$ with an additional elementary $SU(2)$ --- this will allow us a lower compositeness scale but the gauge bosons will be only partially composite. Thus  we need to  introduce an elementary version of the $SU(2)_L$ as well as mixing between these two groups. A straightforward extension of the previous model is to take $F=10$ instead of $F=8$, incorporating two additional $SU(2)$ global symmetries. The decomposition of the dual ``quarks" will then be modified to 

\begin{equation}
\begin{split}
&q= \left(\begin{array}{c} t_n \\ b_n \end{array}\right)_L , \left(\begin{array}{c} c_n \\ s_n \end{array}\right)_L , H_u , H_d^\prime , {\cal H} \\
&\bar q= \left(\begin{array}{c} \nu_e \\ e\end{array}\right)_L , \left(\begin{array}{c} \nu_\mu \\ \mu \end{array}\right)_L ,  \left(\begin{array}{c} \nu_\tau \\ \tau \end{array}\right)_L , \left(\begin{array}{c} u_n \\ d_n \end{array}\right)_L , H_d , H_u^\prime , \bar{\cal H}
\label{Chap3:dualquarks}
\end{split}
\end{equation}
Here ${\cal H, \bar{H}}$ will be bifundamentals under the $SU(2)_{\rm mag}\times SU(2)_{\rm elem}$ group, and the meson coupling will  contain a $P {\cal H \bar{H}}$ interaction term, where $P$ is the singlet in the lower right 2 by 2 corner of the enlarged meson matrix. Together with a tree-level $SU(2)$ invariant mass we can again generate a tadpole for $P$, which will force a VEV for ${\cal H,\bar{H}}$ that breaks the $SU(2)$ groups to the diagonal subgroup. The rest of the considerations from the fully composite model would apply. However, we are forced to add yet more colored states as part of the enlarged meson matrix, and the QCD beta function will increase from the minimum 14 to 17. Due to the presence of the elementary part of the $SU(2)$ gauge symmetry, it is now possible, though not necessary, to also add elementary constituents for the $SU(2)$ doublets as well. We will avoid this option since it would result in an uncomfortably small region of validity of the ``electric'' theory.
This modification of the model corresponds to an RS construction where the $W$, $Z$ and the RH fermions are all partially composites, but the Higgs and the $SU(2)$ doublet fields are fully composite. The amount of compositeness is determined by the relative ratio of the elementary and the ``magnetic" couplings for the lightest fields. The matching of couplings is given by 
\begin{equation}
\frac{1}{g^2}= \frac{1}{g_{\rm comp}^2}+\frac{1}{g_{\rm elem}^2},
\label{couplingmatching}
\end{equation} 
 where $g_{\rm comp}$ is again given by (\ref{magneticmatching}). For $F=10$ we find that the $W$ is about 13 percent composite (the $Z$ also picks up a contribution from the elementary hypercharge gauge boson). If we fix $\Lambda_{\rm el}=10$ TeV we find that (including the effect of running) one needs $F=17$ in order to make the $W$ half composite and half elementary. 
 
We now understand how to produce a partially composite MSSM. However the construction in this section has quite a few more ingredients than what is actually necessary to find a realistic composite MSSM model with no little hierarchy problem. All that is required is that the Higgs and the top quark are composites, and the $W$, $Z$ partially composites. In the next section we present the final most realistic model achieving that.

 %%%%%%%%%%%%%%%%%%%%%%%%%
 %%%%%%%%%%%%%%%%%%%%%%%%%
 \section{The Minimal Model: Partially composite $W$, $Z$, composite top and Higgs, elementary SM fermions}
 \label{sec:minimalpartial}
 \setcounter{equation}{0}
\setcounter{footnote}{0}
 %%%%%%%%%%%%%%%%%%%%%%%%%
 %%%%%%%%%%%%%%%%%%%%%%%%%

The most attractive model is obtained by making only the Higgs and the top composite, the $W$ and $Z$ partially composite, and all other SM fermions elementary. In the Randall-Sundrum picture this would correspond to flat gauge boson profiles, top and Higgs near the IR brane, and the other fermions on the UV brane. There would not be a mixing between elementary and composite fermions for the light SM fields, but rather the Higgs would have a tail on the UV brane giving rise to the elementary fermion masses (this Higgs structure is  a limiting case of the gaugephobic \cite{gaugephobic} model).

 Since we are only asking for the top to be composite, we can lower the number of flavors needed for the duality, which will result in a lot fewer additional colored states eliminating the Landau pole problem for QCD.  We want the global symmetry to contain the full SM gauge group (including an elementary $SU(2)$) and allow the dual ``quarks" to also include a Higgs. Thus we choose $F=6$.  
 The ``electric" theory is 
\beq	
\begin{array}{c|c|cccc}
& SU(4) & SU(6)_1  & SU(6)_2 & U(1)_V & U(1)_R \\
\hline
{\cal Q}& \fund & \overline{\fund}   & {\bf 1} & 1  & {\frac{1}{3}}
\vphantom{\raisebox{3pt}{\asymm}}\\
{\cal \bar Q} & \overline{\fund}   & {\bf 1}  & \overline{\fund}
& -1 & {\frac{1}{3}} \vphantom{\raisebox{3pt}{\asymm}}\\
\end{array}
\label{electricfinal}
\eeq
where the $SU(4)$ is the gauge group while the other groups are the global symmetries.  The usual 
dual theory is
\beq	
\begin{array}{c|c|cccc}
& SU(2)_{\rm mag} & SU(6)_1  & SU(6)_2 & U(1)_V & U(1)_R \\
\hline
q& \fund & \fund & {\bf 1} & 2  & {\frac{2}{3}}
\vphantom{\raisebox{3pt}{\asymm}}\\
\bar q& \overline{\fund}   & {\bf 1}  & \fund
& - 2 & {\frac{2}{3}} \vphantom{\raisebox{3pt}{\asymm}}\\
M& {\bf 1} &  \overline{\fund}   &  \overline{ \fund} & 0 & {\frac{2}{3}}
\vphantom{\raisebox{3pt}{\asymm}}\\
\end{array}
\label{magneticfinal}
\eeq
with the superpotential
\beq
W= y \bar q M q~.
\eeq
As  before the emergent $SU(2)_{\rm mag}$ gauge bosons will partly contain the $W$ and $Z$. The remaining $SU(3)_{c}\times SU(2)_{\rm elem}\times U(1)_Y$ will be embedded in the global symmetries.

In order to identify the top and the Higgses with the dual ``quarks" we use the following embedding of the SM gauge groups in the global symmetry:
\begin{equation}
\begin{split}
&SU(6)_1 \supset SU(3)_{c} \times SU(2)_{\rm elem} \times U(1)_Y \\
&SU(6)_2 \supset SU(3)_X \times SU(2)_{\rm elem} \times U(1)_Y 
\end{split}
\end{equation}
where $SU(3)_X$ is a global $SU(3)$, under which none of the SM fields will eventually transform.
Once again, we will add elementary spectators to cancel anomalies and allow mass terms for all exotic matter fields.
 
The dual ``quarks" can be identified with the SM charged fields as follows:
\begin{equation}
\begin{split}
q&= Q_{3} , {\cal H},  H_d \\
\bar q&= X ,{\cal \bar{H}} ,  H_u 
\end{split}
\end{equation}
where $H_{u,d}$ are the composite Higgses that will eventually give rise to EWSB, while ${\cal H,\bar{H}}$ are bifundamentals under $SU(2)_{\rm mag}\times SU(2)_{\rm elem}$. 
 The meson decomposes as 
\beq
M=\left(\begin{array}{ccc}  V & U & \bar{t} \\ E & G+P & \phi_u \\ R & \phi_d & S \end{array} \right)
\eeq
where $V$ represents three QCD antitriplets, $U$ is a $(\bar{3},2)$, $E$ represents three $SU(2)$ doublets, $G$ is an $SU(2)$ triplet, $\phi_d$ and $\phi_u$ are doublets, $P$ and $S$ are singlets, and $R$ represents three singlets. Note that here we have only added 5 QCD triplets, thus (together with spectators needed to remove these additional triplets) we will have five additional QCD flavors. Above the duality scale one finds that there are only three additional QCD flavors (including spectators).  In this case the QCD $\beta$-function vanishes at one loop and remains perturbative all the way to high scales, and thus no duality cascade of the sort discussed in references~\cite{Strassler,KlebanovStrassler} occurs. Two additional low-scale messengers could be added without spoiling perturbativity below $10^{16}$ GeV. 

The hypercharge assignments for the various fields is fixed by the simple requirement that the $SU(2)_{\rm mag}^2 U(1)_Y$ anomaly cancels, together with the requirement that the dynamical $q M \bar{q}$ superpotential is $U(1)_Y$ invariant, and that the SM fields carry the usual hypercharges. This will fix the hypercharges of the dual quarks to be $\pm {\rm diag} (\frac{1}{6}, \frac{1}{6}, \frac{1}{6}, 0,0,-\frac{1}{2})$ for $q,\bar{q}$. Thus hypercharge can be identified with a combination of two traceless $U(1)$'s inside the the two $SU(6)$'s. From this it is then straightforward to also find the hypercharges of the various components of the meson. The full hypercharge assignments are then
\beq \begin{array}{c|c|c|c|c|c|c|c|c|c|c|c|c|c} 
& Q_3 & {\cal H,\bar{H}} & H_u & H_d& X & V & U & \bar{t} & E & \phi_u & R & \phi_d & G,P,S \\ \hline
Y\vphantom{Y^{Y^{Y^Y}}} & \frac{1}{6} & 0 & \frac{1}{2} & -\frac{1}{2} &  -\frac{1}{6} & 0  & -\frac{1}{6} & -\frac{2}{3} & \frac{1}{6} & -\frac{1}{2} & \frac{2}{3} & \frac{1}{2} & 0 \end{array}~~~.
\eeq
   Note also that after the gauging of the flavor symmetries the only complete singlets are $P$ and $S$, and it is thus natural to assume that these are the only terms that could potentially develop a tadpole term in the superpotential. At this point the magnetic gauge group is anomaly free, however the elementary $SU(2)$ (as well as $SU(3)_{c}$) is not yet. This can be fixed by adding elementary conjugate fields $V', U', \phi_u', R', \phi_d'$ which carry the opposite quantum numbers of $V,U,\phi_u,R,\phi_d$ and thus can get a mass of order the compositeness scale $\Lambda_{\rm el}$, removing these fields from the low-energy spectrum. This leaves us with the additional fields $X, E,P,S$. Note, that $X$ is a singlet under the elementary $SU(2)$, while $E$ is a doublet. The two have exactly the right quantum numbers to transfer the contribution of the $Q_{3}$ to the $SU(2)^2 U(1)$ anomaly from the composite to the elementary sector, while the dynamical superpotential contains the term $ y {\cal H} EX$. One can now add all light fermions (everything except $Q_{3}$ and $\bar{t}$) as elementary fields under the elementary $SU(3)_c\times SU(2)_{\rm elem} \times U(1)_Y$ group, and all anomalies will cancel. 

Baryon number can again be preserved by assigning baryon number 1/3 to $Q_3$ and $-1/3$ to $V,U$ and $\bar{t}$, and zero baryon number to the remaining composites. Similarly all composites have vanishing lepton number. Of course since now most fermions are included as elementary fields we will have no explanation for an accidental low-energy R-parity. What we do find is that the strong dynamics is consistent with the absence of baryon and lepton number violating processes. 

As in the previous sections we will assume a tree-level superpotential (a mass term in the electric theory) which will map into the tadpoles for $S$ and $P$:

\begin{equation}
W\supset y P ({\cal H\bar{H}}-{\cal F}^2)  + y S (H_u H_d -f^2) + y Q_{3} H_u \bar{t}  +  y {\cal H} EX
\end{equation}

The VEV  $\langle \cal H \rangle={\cal F}$ will ensure the breaking of the elementary and composite $SU(2)$'s to the diagonal $SU(2)_L$ subgroup. This will give rise to a massive set of $W'$ and  $Z'$ bosons, the analogs of the lowest Kaluza-Klein  $W'$ and  $Z'$ in RS models. The VEV  $\langle \cal H \rangle$  will also give masses to the fields $E$ and $X$ via the superpotential terms. The masses of these states  are  given by $m_{W'}= \sqrt{g_{\rm el}^2+g_{\rm comp}^2}\, {\cal F} \sim g_{\rm comp}\, {\cal F}$, and $m_{E,X} \sim y \,{\cal F}$.  The masses of the $W'$ and $Z'$ should be around the TeV scale or higher,  while (since $y \sim 1$) $E,X,{\cal H,\bar{H}}$ will also be in this mass range. The matching of the gauge couplings for the low-energy $SU(2)$  is again given by (\ref{couplingmatching}), with the composite coupling evaluated using (\ref{magneticmatching}). For $F=6$ we find $g_{\rm comp}^2\sim 5.2$ and the ``magnetic" $\beta$-function vanishes at one loop (the $SU(4)$ electric theory has no free ``magnetic" phase: $F=6$ is right at the bottom of the conformal window). In this case the $W$ would be about $g^2/g_{\rm comp}^2 \sim 8$ \% composite, since
the low energy effective coupling  will be mostly dominated by the elementary coupling; thus as expected a weakly coupled $W$ and $Z$ is achievable here.  The coupling of the $W'$ and $Z'$ to the elementary quarks and leptons  is also suppressed by $g^2/g_{\rm comp}^2 \sim 8$ \% compared to the SM coupling. If one wants to make the $W$ more composite, one could  increase the number of flavors, while keeping the rest of the construction unchanged. As before, 50 \%  compositeness (with a 10 TeV compositeness scale) would require $F= 17$, and the price would be lots of additional (vector-like) $SU(2)$ doublets around a few hundred GeV. 

\begin{figure}[htb]
\begin{center}
\includegraphics[width=6cm]{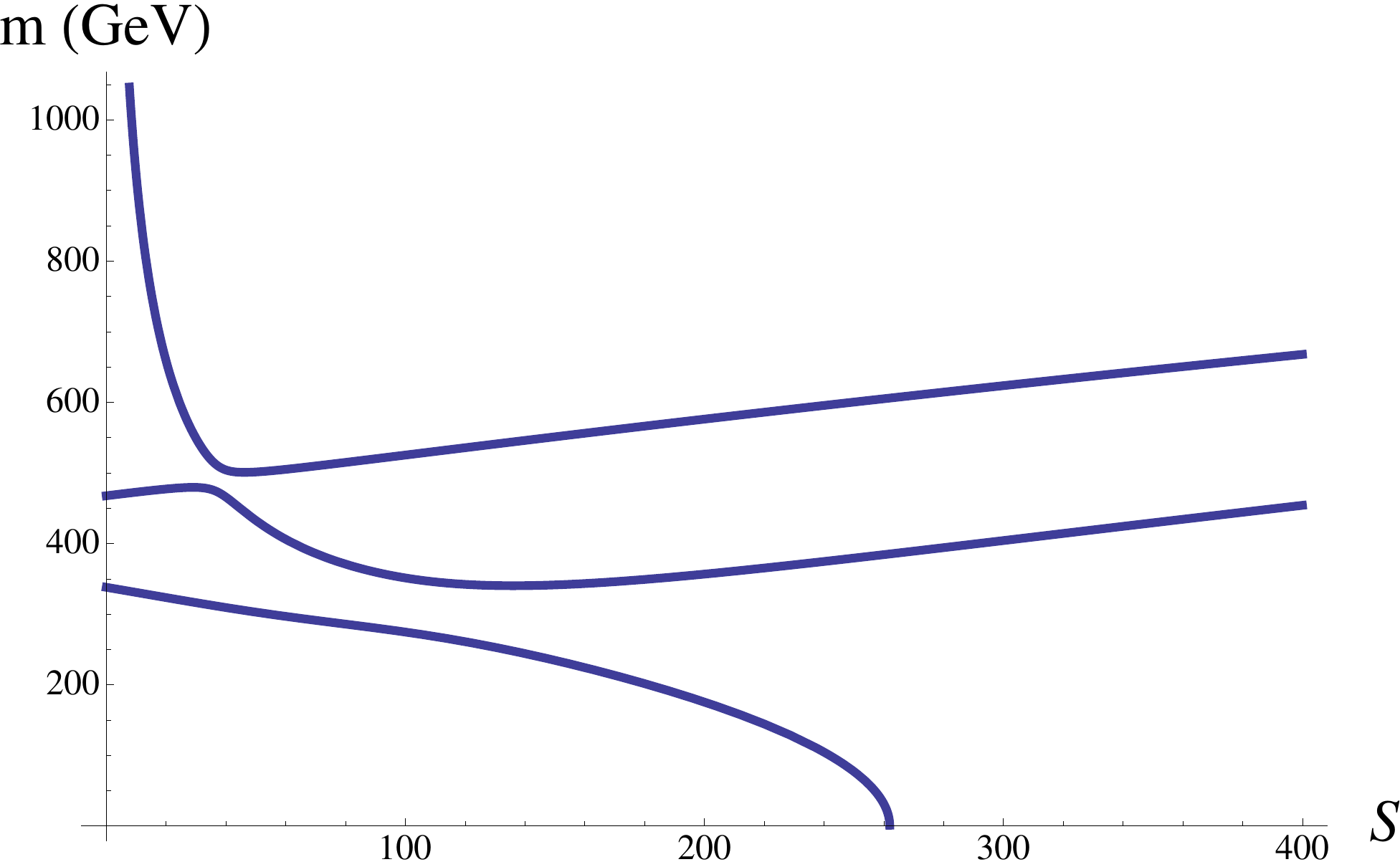} \hspace*{2cm} \includegraphics[width=6cm]{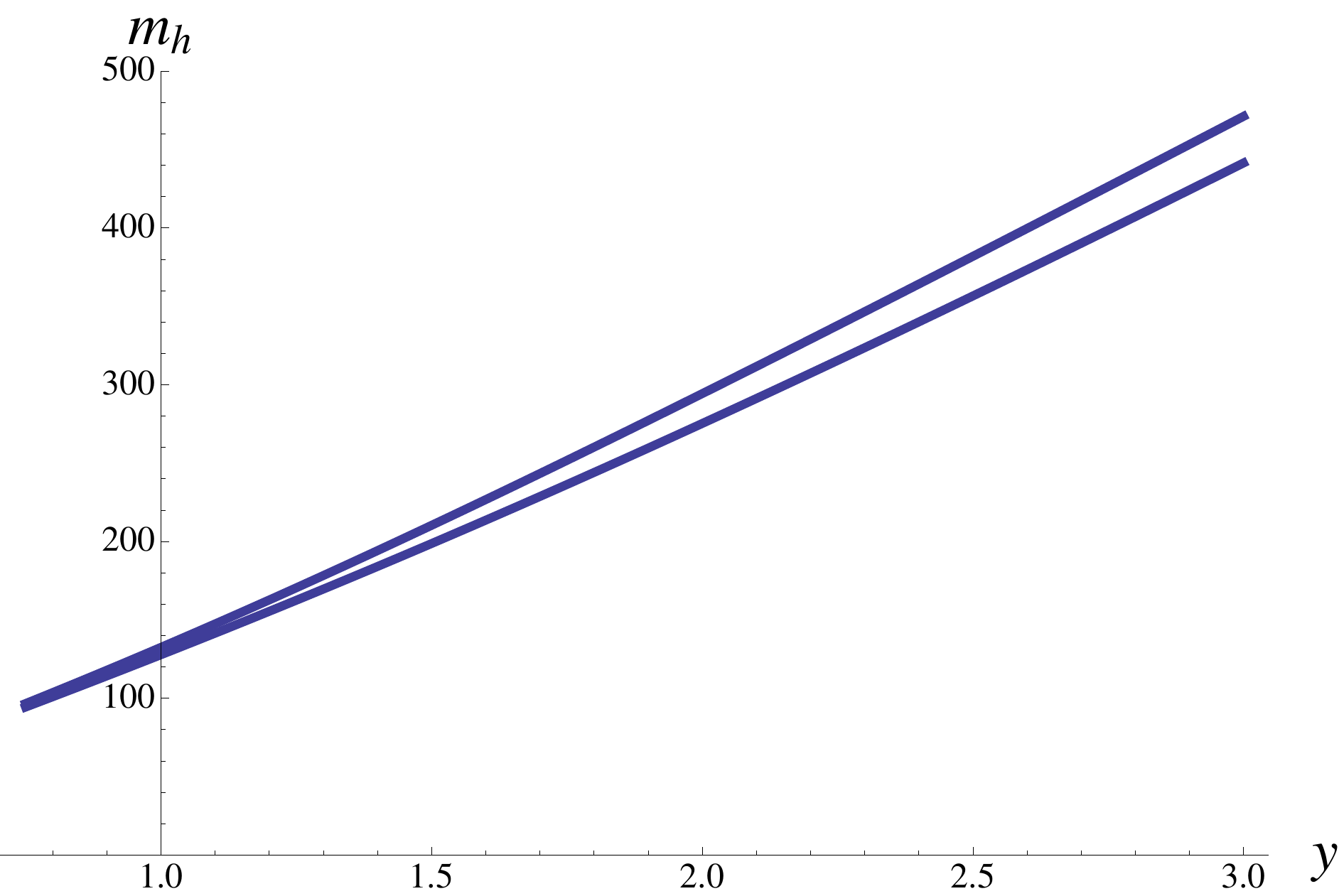}
\end{center}
\caption{Left: The spectrum of CP-even scalars as a function of $\langle S \rangle$ for $\beta = 0.7$,  $f = 200$ GeV, $A = -800$ GeV, and $y = 2$. Right: Upper bound on the mass of lightest CP-even state in GeV, as a function of the Yukawa coupling $y$ for a fixed Higgsino mass $y \,\langle S\rangle = 100$ GeV (upper line) and $y\, \langle S\rangle = 150$ GeV (lower line).}
 \label{fig:spectrum}
\end{figure}

From the point of view of electroweak symmetry breaking and the light fermion masses this final model is basically identical to the usual fat Higgs models \cite{fathiggs1,fathiggs2}. The Higgs potential is given by 
\begin{eqnarray}
&& V= y^2|H_uH_d  -f^2|^2   +y^2|S|^2 (|H_u|^2 +|H_d|^2)  
 +m_S^2 |S|^2 + m_{H_u}^2 |H_u|^2 +m_{H_d}^2 |H_d|^2     \nonumber \\
&&\quad\quad+  A S H_u H_d   +h.c. + 
\frac{g^2+g'^2}{8} (|H_u|^2 -|H_d|^2 )^2  
\end{eqnarray}
where $m_{S,H_u,H_d}^2$ and $A$ are soft supersymmetry breaking parameters, and the last term is the usual MSSM $D$-term. We will use the usual parametrization of the Higgs VEVs $ \langle H_u^0 \rangle = \frac{v}{\sqrt{2}} \sin \beta$, $ \langle H_d^0 \rangle = \frac{v}{\sqrt{2}} \cos \beta$.
In the limit when the $D$-terms  and $A$-terms are neglected the minimum of the potential is given by 
\begin{equation}
\begin{split}
& \tan \beta = \frac{m_{H_d}}{m_{H_u}},  \\
& y^2 v^2 -\frac{4 y^2 f^2}{ \sin 2\beta} + 2 m_{H_u}^2+2 m_{H_d}^2 =0.
\end{split}
\end{equation}
The latter equation is the analog of the usual MSSM fine tuning relation, but as usual in the NMSSM the unknown parameters $y$ and $f$ show up eliminating the need for tuning.
Once the $A$-term is turned on the singlet $S$ will get a VEV of order
\begin{equation}
\langle S \rangle \sim -\frac{A v^2 \sin\beta \cos\beta}{y^2 v^2+m_S^2}
\end{equation}
providing the necessary effective $\mu$-term of the order of the soft breaking scales --- as is usual in NMSSM type models. The full diagonalization of the scalar masses is straightforward numerically. For the electrically neutral CP even sector there will be a three by three mixing matrix among the two neutral Higgses and the CP even singlet scalar. An example spectrum (imposing the correct EWSB minima) is presented in Fig.~\ref{fig:spectrum} together with the upper bound on the lightest Higgs mass. We can see that a Higgs mass of order 400 GeV is possible in this model. 

 Fermion masses for the light fields can be obtained  as in~\cite{fathiggs1,fathiggs2}. The composite Higgses $H_{u,d}$ which obtain VEVs mix with a pair of   Higgses transforming under the elementary group. Recall that among the meson fields we have two doublets under the elementary $SU(2)_{\rm elem}$ group, $\phi_d$ and $\phi_u$, and we have added two additional elementary fields $\phi_d'$ and $\phi_u'$ to pair up with these. $\phi_d'$ has the quantum numbers of an elementary down-type Higgs and $\phi_u'$ that of an elementary up-type Higgs. Thus these fields can have the usual renormalizable Yukawa interactions with the elementary fermions. In addition, we have the composite field $U$ and its elementary conjugate $U'$ which are heavy vectorlike partners of the left handed top-bottom doublet. $U$ mixes with $Q_3$ via the dyanmical superpotential, and $U'$ can have a Yukawa coupling with the elementary Higgses $\phi'_{u,d}$. Taking into account the dynamical superpotential terms after integrating out the additional doublets and the heavy vectorlike quarks the desired effective Yukawa interactions between the composite Higgses $H_{u,d}$ and the elementary fermions will be generated. For example for the quark sector
\begin{eqnarray}
W \supset && y H_u {\cal H} \phi_u + y H_d {\cal \bar{H}} \phi_d + y Q_3 {\cal \bar{H}} U + M_U UU' +M_{\phi_u} \phi_u \phi_u' +M_{\phi_d} \phi_d \phi_d' 
+\nonumber \\&& \sum_{i,j=1,2} \lambda^u_{ij} Q_i \phi_u' \bar{u}_j + \dosum{i=1,2}{j=1,2,3}
  \lambda^d_{ij} Q_i \phi_d' \bar{d}_j +\sum_{j=1,2} \tilde{\lambda}^u_j U' \phi_u' \bar{u}_j +\sum_{j=1,2,3} \tilde{\lambda}^d_j U' \phi_d' \bar{d}_j \nonumber \\ 
\end{eqnarray}
  After integrating out the heavy $\phi_u$, $\phi_u'$, $\phi_d$, $\phi_d'$, $U$, $U'$  fields we find the effective superpotential
\begin{eqnarray}
W_{eff}=&& -\frac{y{\cal F} }{M_{\phi_u} } \sum_{i,j=1,2}  \lambda^u_{ij} Q_i H_u \bar{u}_j - \frac{y{\cal F} }{M_{\phi_d} } \dosum{i=1,2}{j=1,2,3}  \lambda^d_{ij} Q_i H_d\bar{d}_j + \frac{y^2{\cal F}^2 }{M_{\phi_u} M_U} \sum_{j=1,2} \tilde{\lambda}^u_{j} Q_3 H_u \bar{u}_j \nonumber \\ && +\frac{y^2{\cal F}^2 }{M_{\phi_d} M_U}  \sum_{j=1,2,3} \tilde{\lambda}^d_{j} Q_3 H_d \bar{d}_j .
\end{eqnarray}
 This generates the light quark masses, including the CKM mixing matrix. Clearly the down-type mass matrix is the most general one, while for the up-type matrix we get a generic two-by-three matrix connecting the light quarks right handed quarks with all left handed quarks, in addition to the dynamical top quark mass.

Models with composite (fat)  Higgs fields \cite{fathiggs1,fathiggs2} need to address the problem of fitting the electroweak precision measurements.  First of all the $S$ parameter \cite{oblique} tends to grow with the size of the electroweak sector, but (as discussed most recently in \cite{Galloway}) extra contributions to the $S$ parameter from a composite Higgs are suppressed by the square of the VEV over the compositeness scale $v^2/\Lambda^2$, so these contributions are not troublesome here. The precision electroweak fits of the SM prefer a light Higgs, but, as is well known~\cite{PeskinWells}, 
it is possible for the Higgs to be quite heavy and still be consistent with precision electroweak fits due to contributions to $T$. Fat  Higgs models were examined in \cite{Gripaios} where the authors found that Higgs masses as large as 500 GeV were allowed by data.  

Due to the direct product group structure ordinary unification is not expected to happen in this model. However all gauge couplings are perturbative up to high scales (with the exception of the $SU(2)$ composite coupling, for which we go through a Seiberg duality). One may still hope that a unification of the product group sort might happen~\cite{unification}, where for example the elementary $SU(3)_c \times SU(2)_{\rm elem} \times U(1)_Y$ could unify into an $SU(5)$ with the strong coupling providing a threshold correction. This would be analogous to orbifold GUT type unification~\cite{edunification} in extra dimensions. It needs to be seen whether a simple matter content compatible with unification can be found in these theories. It may also be possible to unify the ``electric" $SU(4)$ with $SU(3)_c \times SU(2)_{\rm elem} \times U(1)_Y$.

Finally we comment on flavor changing neutral currents (FCNCs). In the final model presented in this section the only new source for FCNC's (besides the usual MSSM sources like squark masses) are flavor violating couplings of the W' and Z' bosons. These arise because in the gauge basis the 3rd generation LH quarks have a coupling of order $g_{\rm comp}$ to these fields, while the other quarks have a universal coupling of order $g_{\rm elem}$. After rotating to the mass basis there will be flavor-off-diagonal coupling of order $g_{\rm comp} V_{i3} V^*_{3j}$ where $V$ is the CKM matrix. Thus a typical $\Delta F=2$ FCNC 4-Fermi operator will be suppressed by $(V_{i3} V^*_{3j})^2/{\cal F}^2$. Depending on the exact coefficient of these operators (and assuming no new CP violating phase appears) one will find bounds for ${\cal F}$ ranging between 500 GeV and 10 TeV.

%%%%%%%%%%%%%%%%%%%%%%%%%%%%%%%%%%%%%%%%%%%%%%%%%%%%%
%%%%%%%%%%%%%%%%%%%%%%%%%%%%%%%%%%%%%%%%%%%%%%%%%%%%%
\section{Conclusions}
\label{sec:conclusions}
\setcounter{equation}{0}
\setcounter{footnote}{0}
%%%%%%%%%%%%%%%%%%%%%%%%%%%%%%%%%%%%%%%%%%%%%%%%%%%%%

We have presented various possibilities for the $SU(2)_L$ gauge group of the MSSM to arise as a dual ``magnetic" group in the IR. A fully composite $W$ requires a large flavor group to reduce the $SU(2)_L$ gauge coupling. However a very simple model can be found for a partially composite $W$ and $Z$: in this case the composite $SU(2)$ also mixes with an elementary $SU(2)$ reducing the  coupling without a large flavor symmetry. The simplest such model naturally produces an NMSSM-type low energy dynamics with composite Higgs, top and singlet. The little hierarchy problem is absent as in usual fat Higgs models, and the QCD $\beta$-function can be small enough to avoid a duality cascade. 

%%%%%%%%%%%%%%%%%%%%%%%%%%%%%%%%%%%%%%%%%%%%%%%%%%%%%

%%%%%%%%%%%%%%%%%%%%%%%%%%
\section*{Acknowledgements}
%%%%%%%%%%%%%%%%%%%%%%%%%%

We thank Steve Abel, Hsin-Chia Cheng, David Curtin, Andrey Katz, Zohar Komargodski, Markus Luty, Yael Shadmi, Philip Tanedo and Jesse Thaler for useful discussions and comments. The research of C.C. was supported in part
by the NSF grant PHY-0757868. Y.S. is supported in part by NSF grants  PHY-0653656 and PHY-0970173.
J.T. is supported by the US Department of Energy  grant DE-FG02-91ER40674.

%%%%%%%%%%%%%%%%%%%%%%%%%%
\section*{Note Added}
%%%%%%%%%%%%%%%%%%%%%%%%%%
 While we were concluding this work \cite{competition} appeared which explores similar ideas. We  thank Nathaniel Craig and  Daniel Stolarski for correspondence regarding \cite{competition}.

\end{document}